# Electric-Field Control of Magnetic Order: From FeRh to Topological Antiferromagnetic Spintronics

Z. X. Feng[#], H. Yan[#], Z. Q. Liu*

School of Materials Science and Engineering, Beihang University, Beijing 100191, China

[#]These authors contributed equally to this work.

*email: zhiqi@buaa.edu.cn

**Using an electric field instead of an electric current (or a magnetic field) to tailor the electronic properties of magnetic materials is promising for realizing ultralow energy-consuming memory devices because of the suppression of Joule heating, especially when the devices are scaled down to the nanoscale. In this review, we summarize recent results on the giant magnetization and resistivity modulation in a metamagnetic intermetallic alloy – FeRh, which is achieved by electric-field-controlled magnetic phase transitions in multiferroic heterostructures. Furthermore, this approach is extended to topological antiferromagnetic spintronics, which is currently receiving attention in the magnetic society, and the antiferromagnetic order parameter has been able to switch back and forth by a small electric field. In the end, we envision the possibility of manipulating exotic physical phenomena in the emerging topological antiferromagnetic spintronics field via the electric-field approach.**

1. **Background**

Starting from 2002, the total amount of data for the entire society has increased explosively, marking the beginning of the digital age.[1] Since then, more than 90% of the information has been recorded as digital data, most of which are stored on hard disks and in data centers via magnetic recording (schematized in **Figure 1**). The information writing process in magnetic storage requires write heads that consist of small conducting coils and electrical currents to generate magnetic fields for switching the magnetization of ferromagnetic (FM) materials. As a result, even though the switching energy for a tiny bit with a feature size of ~65 nm[2] could be rather small, the Joule heating of the electrical currents in the conducting coils can be significant, which limits the further reduction of energy consumption for information storage.

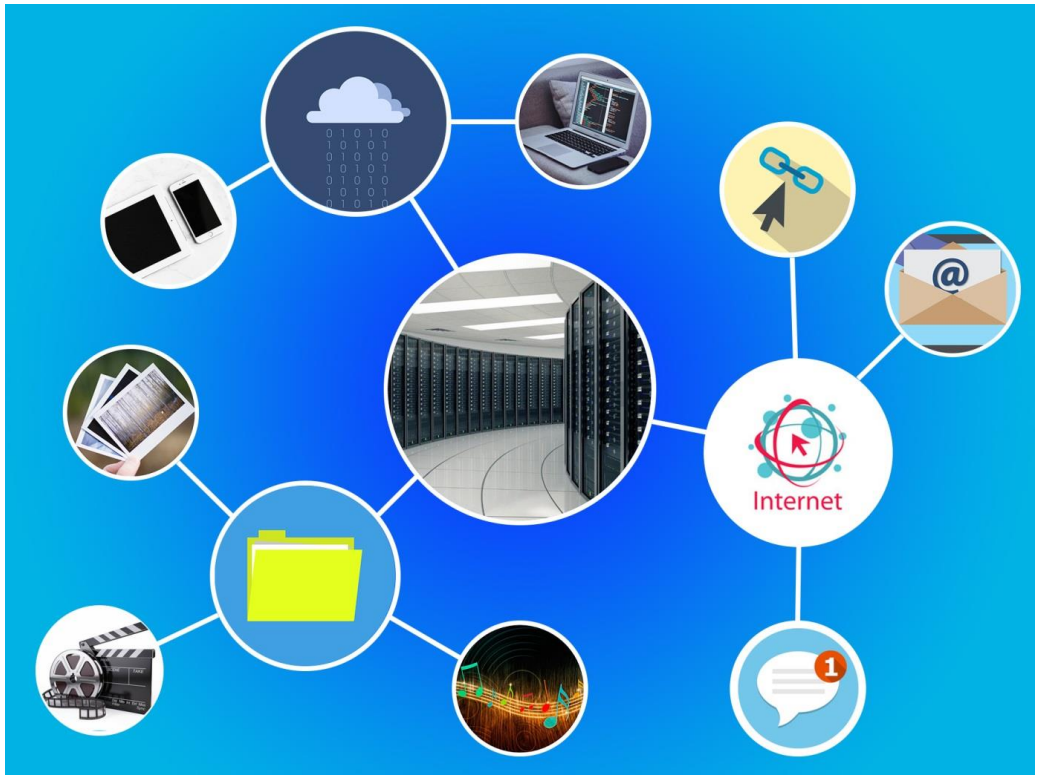

**Figure 1.** Schematic of data storage in the current digital era.

Currently, information technology consumes ~7% of the global electricity, and such energy consumption is expected to increase to ~13% by 2030.[3] Therein, the electricity consumption of data centers would reach ~4.4% of the global electricity consumption, and in particular, the heat dissipation could consume more than 50% of the energy demands of the data centers.[3] Therefore, an alternative information storage approach that is capable of suppressing Joule heating is highly desirable from an energy perspective.

Under this background, electric-field control of magnetism emerges in the field of multiferroics, which is expected to reduce the energy consumption of information storage by several orders of magnitude, to fJ/bit or even aJ/bit.[4–11] In single-phase multiferroic materials, which consists of more than one ferroic order, if there is coupling among different ferroic orders, *e.g.*, the magnetoelectric coupling between the ferroelectric (FE) and FM orders, the magnetism can be conveniently controlled by electrical switching of the ferroelectricity, such as in $YMnO_3$, $BiFeO_3$, and $TbMnO_3$. However, the magnetic order in these materials is typically antiferromagnetic, which is difficult to detect, and the magnetoelectric coupling is rather weak;[12–14] moreover, intrinsic single-phase multiferroic materials are rare because of contradicting symmetry and physical requirements.[15] All these factors prevent single-phase multiferroic materials from practical device applications.

Alternatively, the electric-field control of magnetism can be achieved in heterostructures via other means: 1) in FM/FE composite heterostructures, the magnetism of the FM thin films can be modulated by the piezoelectric strain triggered by electric fields applied onto the ferroelectric substrates;[8] 2) in FM/dielectric composite heterostructures with ultrathin FM thin films, the magnetism can be tailored by the electrostatic doping;[8] 3) in FM/multiferroic composite heterostructures, the magnetism of the FM thin film layers can be varied by electric fields through the interfacial magnetic exchange coupling, such as in $Co_{0.9}Fe_{0.1}$/$BiFeO_3$ heterostructures.[16]

Typically, the modulation of magnetism by electric fields in multiferroic heterostructures results in variations in the magnetization, coercivity fields or magnetic anisotropy of the ferromagnetic materials. In addition, the key scientific issue of controlling magnetic properties using an electric field has been carefully elaborated in previous review articles.[17,18] Among them, the change of magnetization ($\Delta M$) upon external electric fields ($\Delta E$) yields the simplified magnetoelectric coupling coefficient (strictly speaking, the magnetoelectric coupling coefficient should be described as a tensor; please refer to the previous review articles)[17,18] $\alpha = \mu_0 \Delta M/\Delta E$, where $\mu_0$ is the permittivity of free space. Compared with single-phase multiferroics, $\alpha$ in multiferroic heterostructures can be significantly greater, $e.g.$, it reaches $1.08 \times 10^{-7}$ s m$^{-1}$ in $Fe_3O_4$/$Pb(Mg_{1/3}Nb_{2/3})_{0.7}Ti_{0.3}O_3$ ($Fe_3O_4$/PMN-PT) heterostructures,[19] $2.3 \times 10^{-7}$ s m$^{-1}$ in $La_{0.67}Sr_{0.33}MnO_3$/$BaTiO_3$ heterostructures[20] and $Co_{40}Fe_{40}B_{20}$/PMN-PT heterostructures,[21] while it is only $10^{-12}$ s m$^{-1}$ in $Cr_2O_3$,[22] $10^{-10}$ s m$^{-1}$ in $TbMnO_3$[12] and $10^{-9}$ s·m$^{-1}$ in $Ni_3B_7O_{13}I$.[23]

## 2. Introduction to FeRh

In 2014, the magnetoelectric coupling coefficient was further enhanced to a new record of ~ $1.6 \times 10^{-5}$ s m$^{-1}$ in multiferroic FeRh/$BaTiO_3$ heterostructures by Cherifi $et\ al.$,[24] where $CsCl$-type (group $Pm3m$)[25] FeRh is a metamagnetic (metamagnetism typically means a dramatic increase in the magnetization of a material under an external magnetic field) intermetallic alloy and exhibits a first-order magnetic phase transition (MPT) from a low-temperature antiferromagnetic (AFM) phase to a high-temperature FM phase.[25–30] Accompanied by the MPT, the resistivity and the lattice constant are changed as well (**Figure 2**). It is therefore a material system with strong correlations among lattice, magnetization and electrical transport properties. As a result, such a system is an ideal candidate for electric field control of magnetism and resistivity via piezoelectric strain. The reason is that the electric-field-induced piezoelectric strain in ferroelectric oxide substrates can conveniently modulate the lattice degree of freedom of any thin films grown on top of them and accordingly change the magnetic phase and electrical resistance. The unique first-order AFM-FM transition has the potential of enabling the full magnetic phase transformation induced by electric fields, which thus could generate giant magnetization changes compared with previous multiferroic heterostructures, which typically exhibit partial magnetization modulation or magnetic anisotropy changes under electric fields. This finding is also the reason that the largest magnetoelectric coupling coefficient until now has been achieved in FeRh-based multiferroic heterostructures.

The phase transition temperature in FeRh strongly depends on the chemical ratio of Fe and Rh,[31] the strain state,[29,32,33] magnetic fields,[24,33,34] chemical doping,[33–37] and ion irradiation.[38,39] Nevertheless, the physical origin for the metamagnetic transition in FeRh is still under intensive debate, and a few mechanisms have been proposed, such as lattice expansion-induced exchange inversion,[40] electronic entropy,[41] instability of Rh moment,[42] spin wave excitation[43] and magnetic excitations.[44–46]

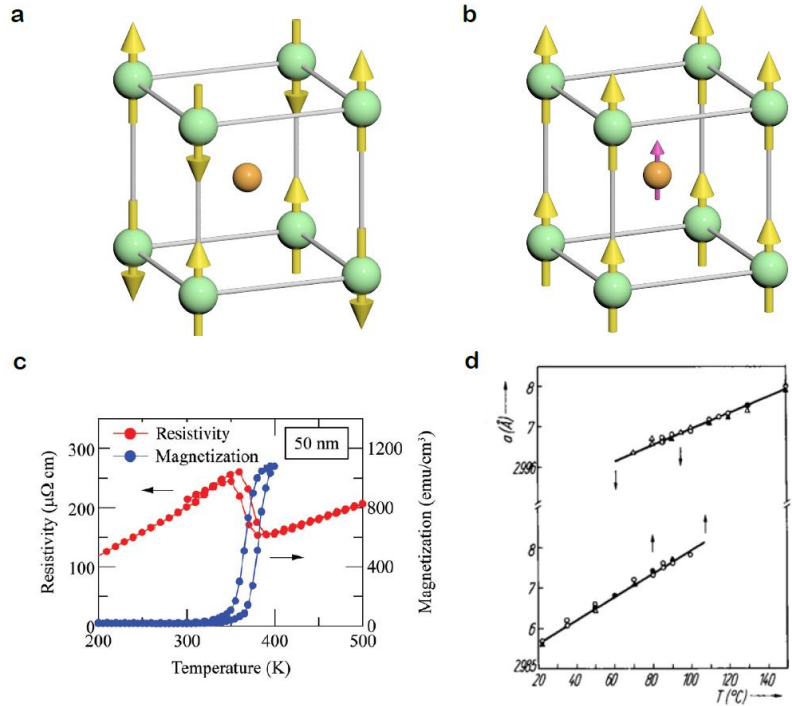

**Figure 2.** Structural, electrical and magnetic properties of FeRh. a,b) Schematics of the low-temperature AFM phase and the high-temperature FM phase of intermetallic FeRh, respectively. c) Resistivity and magnetization of a 50-nm-thick FeRh film. Reproduced with permission.[47] Copyright 2011, AIP Publishing. d) Temperature-dependent lattice constant of bulk FeRh. Reproduced with permission.[48] Copyright 1967, John Wiley and Sons.

Despite the controversial origin of the temperature-driven metamagnetic transition, FeRh has shown great potential in heat-assisted information storage as well as magnetocaloric cooling applications. For example, Thiele *et al.* demonstrated that FeRh can be used as an auxiliary layer for thermally assisted magnetic recording media.[49] Similarly, it was found that the AFM to FM transition can be excited by ultrafast (500 fs – 30 ps) laser pulses due to the heating effect, which could be useful for optical information writing.[50,51] Moreover, FeRh exhibits a giant magnetocaloric effect,[52–57] which leads to a large effective refrigerant capacity of 324.42 J·kg$^{-1}$.[53]

In addition, the successful electrical switching of the MPT in FeRh by Cherifi *et al.* created exciting opportunities for utilizing FeRh in low energy-consuming memory devices.[24] It was first achieved in multiferroic FeRh/BaTiO$_3$ heterostructures that consisted of epitaxial FeRh thin films and ferroelectric BaTiO$_3$ single-crystal substrates. Because the metamagnetic phase transition temperature of bulk FeRh is rather sensitive to pressure,[29] equivalently, it can be largely modulated by the strain in FeRh thin films. On the other hand, ferroelectric materials are capable of effectively generating long-range, dynamic and reversible piezoelectric strain when they are subject to electric fields.[58,59] While the FeRh films are epitaxially grown on the BaTiO$_3$ substrates, the electric-field-induced piezoelectric strain in BaTiO$_3$ can be almost 100% transferred into the FeRh thin films through interface mediation, which results in a highly effective modulation.[24] It turns out that such pioneering work has rapidly excited a surge of studies on electric-field control of magnetic order in FeRh since 2014.

In the following, we will first revisit epitaxial growth of FeRh thin films on oxides, and then, we summarize the research progress on the electric-field manipulation of its magnetic order. Subsequently, we will briefly introduce the cutting-edge topological AFM spintronics and overview our recent results on controlling the AFM order parameter via the electric-field approach. Finally, we will envision the possibility of controlling other exotic topological effects, such as magnetic Weyl fermions in antiferromagnets by electric fields.

### 3. Epitaxial growth of FeRh on oxides

Overall, epitaxially growing intermetallic alloys on oxide has been a challenging task due to several factors:[60] 1) epitaxy naturally requires high temperatures for substrates during growth because the atoms require sufficient thermal energy to move to equilibrium lattice sites, which, on the other hand, could easily result in the oxidation of intermetallic alloy films or new alloying at interfaces; 2) since the growth of intermetallic alloys films are performed in an oxygen-deficient atmosphere and these materials usually contain chemically reactive metal atoms, they could easily capture the oxygen atoms on the surfaces of oxide substrates to form oxidation even though the growth is performed at room temperature;[61] 3) as a result of the large difference in surface energy between intermetallic alloys and oxides, there is the wetting issue, which typically leads to balling of high-temperature-grown films and could break epitaxy; 4) in terms of lattice constants, intermetallic alloys differ from most oxides (3.7-4.3 Å),[62] which could prevent thin films from epitaxial growth.

Therefore, adequate oxide substrates should be chemically stable at high temperatures and have as little lattice mismatch as possible. For example, $SrTiO_3$ might not be an ideal substrate for growing intermetallic alloys at high temperatures. The reason is that it has a high oxygen diffusion constant and can lose oxygen in an oxygen-deficient environment,[63–65] which leads to interfacial oxidation. In contrast, MgO is an excellent substrate for the epitaxial growth of FeRh.[34,66,67] On the one hand, it is highly stable at high temperatures. On the other hand, it is cubic with a lattice constant of 4.216 Å and nicely matches the lattice of FeRh ($a$ = 2.99 Å) considering an in-plane lattice rotation by 45° (2.99×$\sqrt{2}$ = 4.228 Å). As shown in **Figure 3**a & b, the epitaxial growth of FeRh on MgO is evidenced by the high-quality interface and the X-ray diffraction spectra.

The interfaces between FeRh films and oxide substrates could play an important role in the magnetic properties of FeRh due to possible strain and intermixing effects. The epitaxial strain generated at the interfaces between MgO substrates and FeRh films was found to largely affect the magnetic phase transition of FeRh.[68] It is also noticed that although bulk FeRh is AFM at room temperature, all the FeRh films grown on BTO substrates exhibit a relatively large magnetization at room temperature,[24,69,70] which could be well due to the interface effects. The sensitivity of the FeRh phase transition temperature to epitaxial strain could benefit the realization of the room temperature magnetic phase transition without doping, which is favorable for memory device applications. In addition, the intermixing between metal films and oxides can be useful to realize exotic magnetoelectric coupling as well. For example, in an Fe/BTO multiferroic system, the polarization (or ion displacement) in BTO controls the exchange coupling constants of an ultrathin intermixing FeO$_x$ layer at the interface, which leads to electrical switching of the ferromagnetism between "on" and "off".[71]

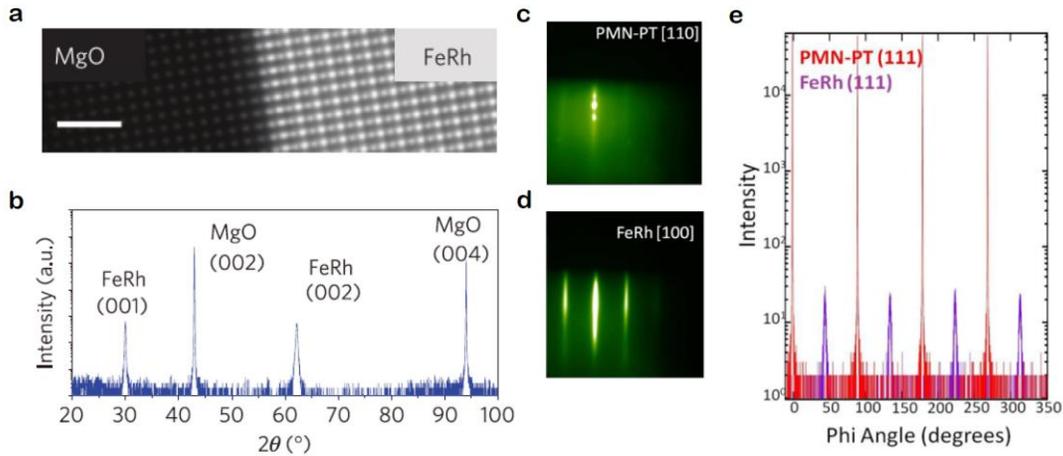

**Figure 3.** Epitaxial growth of FeRh films on oxides. a,b) Cross-section transmission electron microscopy image and X-ray diffraction spectrum of an FeRh/MgO heterostructure, respectively. Reproduced with permission.[66] Copyright 2014, Springer Nature. c,d) Reflection high-energy electron diffraction patterns of a PMN-PT substrate along its [110] crystallographic direction and the FeRh film grown on it via molecular beam epitaxy, respectively. e) 360° Phi scans around the (111) peaks of PMN-PT and FeRh for the epitaxial FeRh/PMN-PT heterostructure, as shown in (c) and (d).

Regarding the growth technique of FeRh thin films, sputtering has been extensively utilized in previous studies,[3,8,13,22,32,35,40,46,59,67,70,72–77] where the growth temperature ranges from 250 to 630 °C, and the Ar pressure is 3-6.5 mTorr. To enhance the chemical ordering of the FeRh thin films, vacuum annealing at higher temperatures that range from 400 to 750 °C can be further performed after deposition. The sputtering power source can be either d.c. or a.c. It is worthwhile to note that the a.c. sputtering yields a relatively slower growth rate compared with the d.c. technique, and a larger target-substrate distance could be useful in achieving high quality interfaces.[69] In addition, molecular beam epitaxy has been used for fabricating epitaxial FeRh thin films,[78] *e.g.*, FeRh/PMN-PT heterostructures (**Figure 3**c, d & e).

## 4. Electric-field control of the magnetic order in FeRh

The first report on the magnetic intermetallic alloy FeRh appeared in 1938, where Fallot prepared ferromagnetic Fe alloys with Pt, Ir, Os, Ru, Rh, and Pd.[26] One year later, Fallot and Hocart discovered a sharp increase in the magnetization of approximately equiatomic FeRh alloys when the temperature was increased to ~350 K.[79] They also observed a temperature hysteresis associated with the magnetization change, which suggests a first-order transition.[27] Later, in 1961, de Bergevin and Muldawer demonstrated that the transition induced by increasing the temperature does not affect its *CsCl*-type phase but results in a uniform volume expansion of ~1%.[80] Furthermore, they proved that such a transition is a first-order AFM-FM transition via neutron diffraction measurements.[81] In 1962, Kouvel and Hartelius showed that a large resistivity drop (more than 40%) occurs accompanied by the AFM-FM transition.[27] Recently, de Vries *et al*. performed detailed Hall measurements and found that the carrier density of the FM phase is one order of magnitude higher than the AFM phase, which implies significant changes in the densities of the states and the band structure.[82] Therefore, a near equiatomic FeRh alloy is a material system with direct correlations among lattice, magnetization and electrical transport properties.

### 4.1 Giant magnetization modulation.

In 2014, Cherifi *et al*.[24] epitaxially grew 20-nm-thick FeRh thin films on (001)-oriented $BaTiO_3$ (BTO) single-crystal substrates and investigated the effect of electric fields on the MPT of FeRh. They found that the piezoelectric strain triggered by a small electric field of 0.4 kV cm$^{-1}$ enhanced the MPT temperature by ~25 K. More importantly, at approximately the MPT point, the in-plane compressive strain reversed its original FM phase into the AFM phase (**Figure 4**a and b). Accordingly, an enormous magnetization modulation of ~270 emu·cm$^{-3}$ was achieved

(**Figure 4**c), which results in the (until now) largest magnetoelectric coupling coefficient $\alpha$ ~$1.6\times10^{-5}$ s m$^{-1}$. In 2018, Xie *et al.* analyzed the FM hysteresis loops of 30-nm-thick FeRh films grown on FE PMN-PT substrates as a function of the electric field and temperature. They obtained a very large coercivity field change of ~260% controlled by the electric field around the MPT temperature (**Figure 4**d),[77] which is consistent with the piezoelectric strain-induced magnetic phase change in FeRh. The intrinsic origin of the giant electric modulation of the magnetic properties in FeRh is that around the phase transition, multiple phases possess comparable Gibbs free energies, and thus, the material system is rather sensitive to external stimuli such as magnetic fields, strain, and possibly optical excitations.[69]

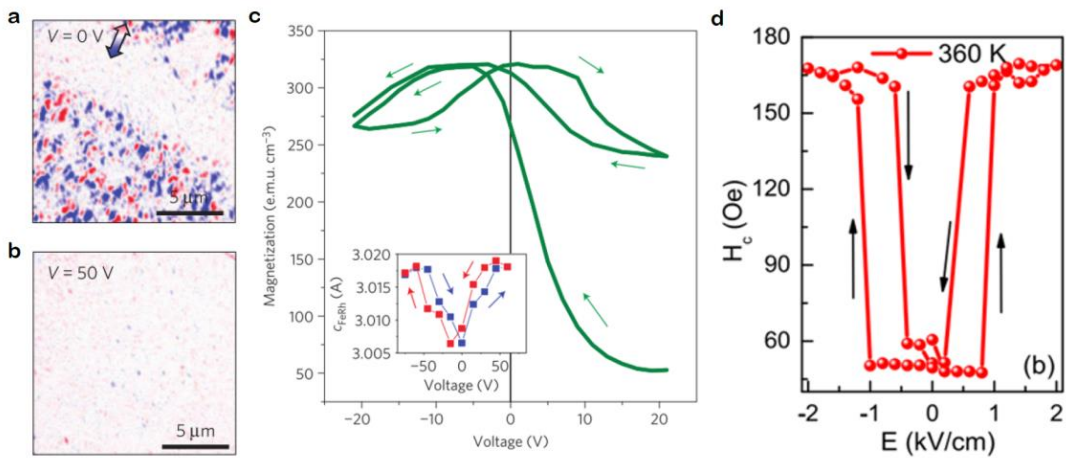

**Figure 4.** Electric-field control of the magnetic properties of FeRh. a,b) X-ray circular dichroism-photoemission electron microscopy of a 20-nm-thick FeRh film grown on a BTO substrate with zero voltage and 50 V applied to the BTO substrate, respectively. Reproduced with permission.[24] Copyright 2014, Springer Nature. c) Voltage-dependent magnetization in the FeRh/BTO heterostructure. Inset: Out-of-plane lattice constant of the FeRh film as a function of the voltage. Reproduced with permission.[24] Copyright 2014, Springer Nature. d) Electric-field-dependent coercivity field of a 30-nm-thick FeRh film grown on a PMN-PT substrate. Reproduced under the terms of the Creative Commons Attribution (CC BY) license.[77] Copyright 2018, AIP Publishing.

### 4.2 Giant electroresistance modulation.

Soon after the report of the giant magnetization modulation by Cherifi *et al.*,[24] Lee *et al.*

fabricated textured 50-nm-thick FeRh thin films with a preferred (001) orientation on PMN-PT substrates, and they realized a maximal resistivity modulation of ~8% via the piezoelectric strain at 368 K (**Figure 5**a),[78] which was demonstrated to originate from the partial magnetic phase conversion between the FM and AFM phases. This work confirms that the piezoelectric strain can effectively induce an isothermal magnetic phase transition in FeRh thin films as discovered by the previous work.[24] Furthermore, it realizes a nonvolatile resistive memory (**Figure 5**b) based on a single intermetallic layer, which could facilitate applications due to the simplicity of the memory structure and the convenience of the information readout.

An 8% electroresistance modulation is already remarkably large for a metallic system with a carrier density of $10^{22}$-$10^{23}$ cm$^{-3}$. The reason is that the Thomas-Fermi screening length of such a highly conductive system is ~2 Å,[69] and thus, the electrostatic modulation via carrier injection/depletion would be less likely to work for 50-nm-thick films. Even though the ionic liquid gating could yield a large carrier density modulation of ~$10^{15}$ cm$^{-2}$, only a 2.3% resistivity modulation was obtained in a thin Au film at 220 K.[83]

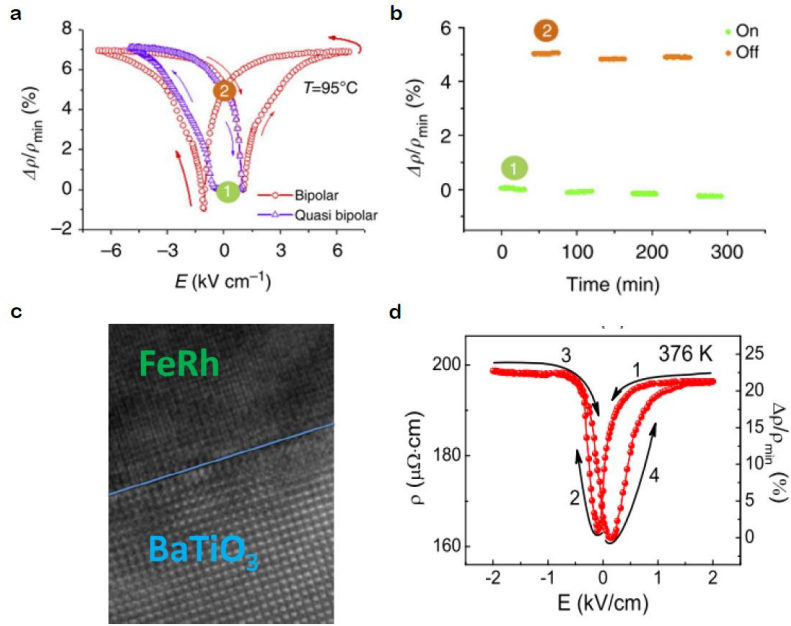

**Figure 5.** Giant electroresistance. a) Bipolar and unipolar electroresistivity of a 50-nm-thick FeRh/PMN-PT heterostructure at 368 K.[78] b) Nonvolatile high and low resistance states achieved by electric pulses of -6.7 and +1.8 kV cm$^{-1}$, respectively.[78] c) Cross-section image of a 35-nm-thick FeRh/BTO heterostructure. d) 22% electroresistance modulation in the FeRh/BTO heterostructure shown in (c).[69]

PMN-PT single crystals are not chemically stable at high temperatures because Pb is volatile. Therefore, possible interdiffusion between FeRh and Pb that could further result in the oxidation of Fe and secondary alloys would occur, which would lead to the poor crystallinity and polycrystalline texture of FeRh films. To avoid this issue, FeRh films were grown on BTO single-crystal substrates at similar conditions as in Liu *et al*.[69] It was determined that both the FeRh crystallinity and the interfaces (**Figure 5**c) between the FeRh films and the substrates are largely improved. As a result, an ~22% electroresistance modulation was achieved at 376 K (**Figure 5**d). This value is comparable to the highest room temperature giant magnetoresistance ratio of ~21% achieved in multilayer structures in 1995,[84] which is eight years after the discovery of the giant magnetoresistance effect.[85,86] Therefore, such a large electroresistance in FeRh was coined as giant electroresistance for metals.[69]

Other studies also confirm that the magnetic and electrical properties of FeRh are sensitive to strain, *e.g.*, Chen *et al.*'s work[70], which examines the effect of BTO structural phase transitions on the MPT of FeRh, and Xie *et al.*'s work[73], which investigates the effect of epitaxial strain on the magnetic and transport behavior of FeRh films. Moreover, Hu *et al.*[87] achieved a remarkable electric-field-controlled magnetocaloric effect in FeRh$_{0.96}$Pd$_{0.04}$/PMN-PT heterostructures based on the magnetic phase conversion triggered by the piezoelectric strain, which extends the electric-field approach to refrigeration applications.

**4.3 Dominant stimulus for an isothermal magnetic phase transition.**

Usually, phase transitions result from the competition of free energies of different phases. Whichever phase possesses a lower free energy under external stimuli, it would become more stable. Around the phase transition temperature, multiple phases are comparable in terms of the free energy, and thus, the phase transformation among them could be triggered by even subtle energetic excitations.[69] On the other hand, the free energies of different magnetic phases of FeRh were found to be rather sensitive to pressure and, thus, to lattice variation.[29] As a result, isothermal magnetic phase transitions in FeRh around its phase transition temperature have been achieved by small electric fields.

However, it should be asked what is the dominant stimulus in the isothermal magnetic phase transition? As is well known, the conventional MPT in FeRh upon varying the temperature is closely associated with the lattice volume change (~1%). However, the dominant driving factor for the newly realized isothermal MPT in FeRh controlled by the electric-field/piezoelectric strain remains an open question. Accordingly, Liu *et al.* measured the lattice change in a 35-nm-thick FeRh/BTO heterostructure that exhibits a full electric-field-controlled MPT as a result of the enhanced film quality, and they revealed that the predominant factor for isothermal MPT is not the lattice volume expansion but is the lattice distortion – the tetragonality (*c*/*a*) change (**Table 1**).[69] This finding provides a new driving mechanism for the MPT in FeRh.

Table 1. Tetragonality and volume changes of FeRh films for different MPTs

| Different magnetic phase transition (MPT) in FeRh | Tetragonality change (%) | Volume change (%) |
|---|---|---|
| *T*-driven MPT in FeRh/BTO | 0.12% | 1.13% (similar to bulk) |
| Isothermal strain-induced MPT in FeRh/BTO | 0.72% | 0.17% |
| Isothermal strain-induced partial MPT in FeRh/PMN-PT | 0.17% | 0.05% |

### 4.4 Super low energy consumption.

Because FE oxides are highly insulating, an electric field leads to only a negligible current (typically below nA) during electrical switching, and thus, the Joule heating is largely suppressed. Considering slightly structure made of an FeRh/BTO heterostructure with footprint dimensions of 100×100×100 nm$^3$, the switching energy for writing slightly is $E_{Switching} = \frac{1}{2}P_S V_{Switching} S_{Bit} = 0.5 \times 26$ μC cm$^{-2}$ × 0.4 kV cm$^{-1}$ × 100 nm × 100$^2$ nm$^2$ ≈ 5.2×10$^{-18}$ J = 5.2 aJ, where $P_S$ is the saturation polarization of BTO, $V_{Switching}$ is the polarization switching voltage of BTO during information writing using the piezoelectric strain, and $S_{Bit}$ stands for the device area for a bit. This switching energy is more than 7 orders of magnitude lower than the energy consumption of the contemporary MRAM devices, which is on the order of 10$^{-10}$ J/bit.

### 4.5 Challenges for memory applications

However, to realize the information storage technology of high-density and super-low (aJ) power costs using these FM/FE heterostructures, there are several substantial challenges to overcome. First, when bulk FE oxides are integrated onto Si substrates as thin films, the electric-field-induced piezoelectric strain would be largely discounted due to the clamping effect from the film/substrate interfaces; second, compared with bulk materials, the coercivity fields of FE oxides could be largely increased in thin films due to the enhanced density of defects, which would increase the polarization switching voltage and thus the power consumption in small devices; third, when the memory devices are scaled down to the nanoscale (< 100 nm), the FE oxides could likely be single-domain, which could be unfavorable to the ferroelastic strain generation.

Hence, more studies on solving these issues are needed to enable practical application of these devices. For example, patterning FE oxide films into nanoscale islands could be useful to relieve the clamping effect; optimizing the FE oxide film growth via controlling the growth parameters to suppress the density of the defects would be desired; and tailoring FE domains with exotic features could be favorable in realizing multiple-domain structures, even at the nanoscale.[88]

### 4.5 "Caution" for bulk FE substrates

In the above, we have summarized the electric-field-control of magnetic, electrical and caloric properties of FeRh in FeRh/FE heterostructures. Although FeRh is our key functional material here, bulk FE substrates play an essential role in converting the electric field into the strain. Under cyclic electric fields, bulk FE materials could crack due to the internal microscopic stress accumulation at the domain boundaries between freely switchable domains and other domains pinned by surface defects.[89,90] As a result, reversible opening and closing of nanoscale cracks can be observed, *e.g.*, in PMN-PT (**Figure 6**a and b). These cracks are typically induced by repeating bipolar switching and are detrimental for device applications. However, they could be avoided by unipolar switching in FE materials. On the other hand, once the nanoscale cracks appear in the FE substrates, they could result in an enormous resistance change[90] and even possibly affect the magnetic properties[91] of the thin films grown on top of them. Therefore, one should exercise special caution when using FE oxide substrates for electric-field modulation of physical properties of thin films.

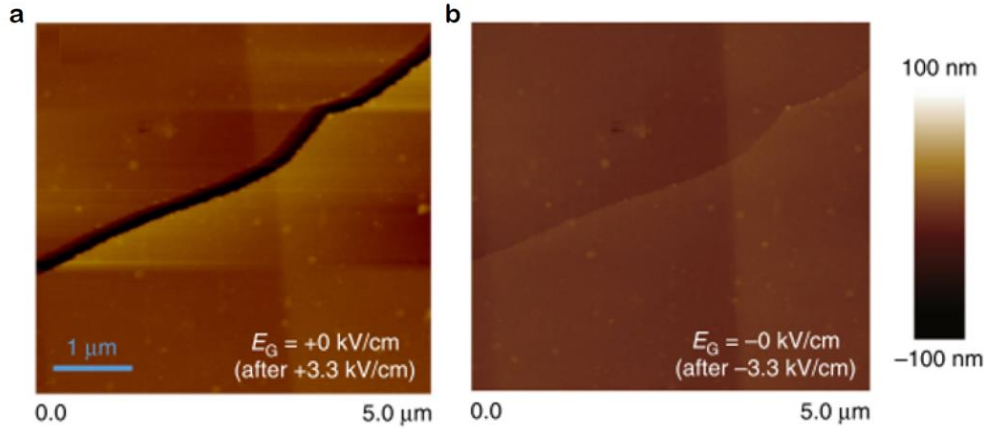

**Figure 6.** Reversible crack driven by an electric field in a 35-nm-thick MnPt/PMN-PT heterostructure. Atomic force microscopy images (5 × 5 µm$^2$) of a single crack in the MnPt film after scanning the gate electric field $E_G$ a) from +3.3 kV cm$^{-1}$ to 0 kV cm$^{-1}$ and b) from −3.3 kV cm$^{-1}$ to 0 kV cm$^{-1}$.[90]

**4.6 Giant electrochemical modulation.**

In addition to the piezoelectric control of the magnetic order in FeRh/FE heterostructures, the ionic-liquid-gated electric double layer structure has been utilized to modulate the magnetic phase of FeRh.[72] As a consequence of the electrostatic carrier injection/depletion, they successfully modulated the MPT temperature for a 5-nm-thick FeRh film grown on MgO by more than 80 K (**Figure 7**), which is much larger than the piezoelectric strain modulation (~25 K) realized by FE oxide substrates. Although the ionic liquid approach is currently not favorable for fast memory device applications because it usually takes half hour to get stable modulation performance upon every switching, the giant electrochemical modulation demonstrated by Jiang et al.[72] implies that such a method could be promising for creating exotic physical phenomena in FeRh and other metallic systems.

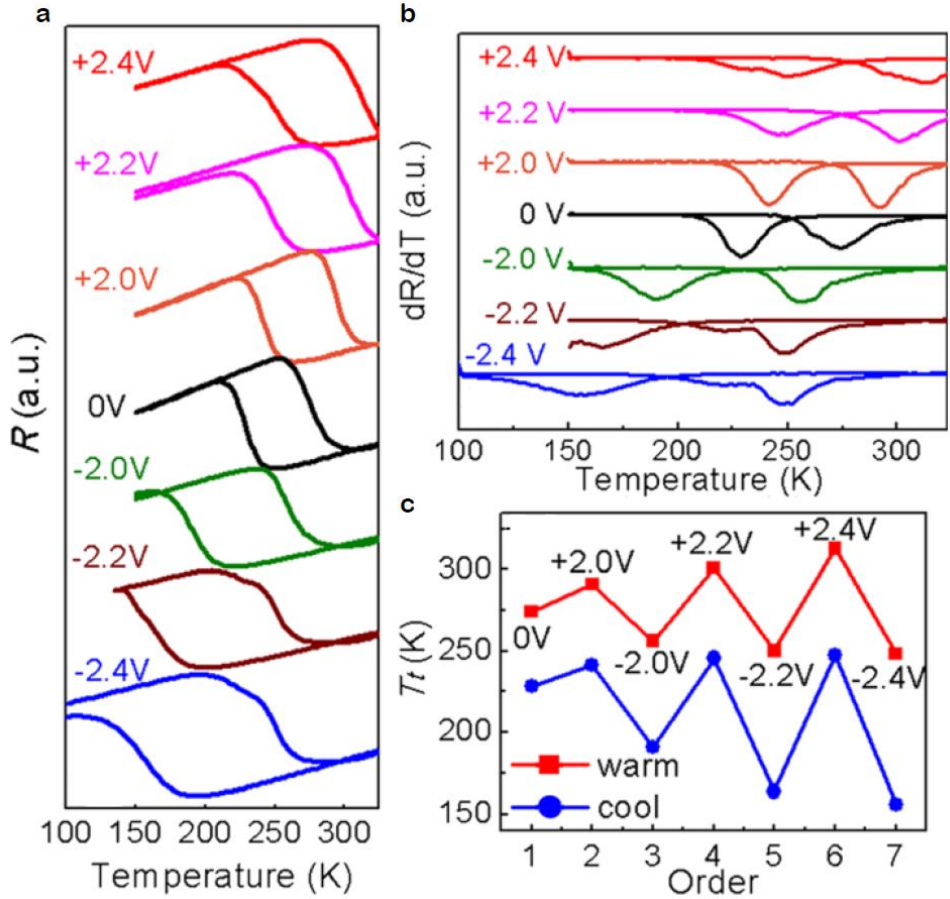

**Figure 7.** Electrochemical modulation of the FeRh phase transition. a) Temperature dependent resistance and b) the resistance variation upon temperature cycling ($dR/dT$) at different gate voltages $V_G$ with fields of 70 kOe. c) $V_G$ dependence of the transition temperature of FeRh for both warming and cooling processes. Reproduced with permission.[72] Copyright 2016, AIP Publishing.

## 5. Introduction to the topological AFM spintronics

AFM spintronics is an emerging field that focuses on manipulating the AFM order parameter – spin axis using electric, optical and other means of control for information technology devices.[92,93] The pioneering theory on the AFM spintronics was proposed in 2006,[94] *i.e.*, spin torques and giant magnetoresistance could be realized in AFM metals similar to the spintronics in FM materials. More importantly, it was emphasized that the critical current for switching the AFM order parameter orientation via spin torques can be smaller than that for switching the FM magnetization because of the absence of shape anisotropy in an AFM metal and because spin torques can act through the entire volume of an antiferromagnet.[94] Later, in 2011, Park *et al*. successfully demonstrated an antiferromagnet-based spin-valve-like tunnel junction,[95] where the rotation of the spin axis in AFM MnIr enabled by the exchange spring effect between MnIr and FM NiFe affects the perpendicular tunneling of electrons. This finding marks the very start of experimental AFM spintronics.

Compared with FM materials, AFM materials have no stray magnetic fields and thus are insensitive to external magnetic fields. On the other hand, owing to the canting of the magnetic sublattices, the zero-field AFM resonance frequency involves the AFM exchange field $H_{ex}$ in addition to the anisotropy field $H_a$, which solely determines the zero-field resonance frequency of the FM materials and usually reaches THz[96], which is much higher than the FM resonance frequency (~GHz). Hence, memory devices built on AFM materials could be capable of resisting magnetic fields and could have a faster information writing speed of ~ps.

Nevertheless, the effective manipulation of the AFM spin axis is much more difficult than the control of FM magnetization. In 2014, Marti *et al*.[66] applied a magnetic field of 9 T during cooling an FeRh film through its FM-AFM transition and found that the spin axis of the low-temperature AFM phase is perpendicular to the cooling field. Accordingly, the orientation of the AFM spin axis in FeRh can be controlled by the direction of the cooling field applied above the MPT temperature of FeRh. Due to the relativistic spin-orbit coupling, anisotropic magnetoresistance (AMR) exists for AFM materials as well. Similar to the AMR effect in FM materials, the resistance is higher when the AFM spin axis is parallel to the measuring current and is lower when the AFM spin axis is transverse to the current.[66] As a result, they demonstrated the concept of an AFM memory resistor. In 2016, Wadley *et al*.[97] realized the electrical switching of the AFM order parameter in the AFM semiconductor CuMnAs via the

Néel spin-orbit torque (SOT) and different resistance states originating from the AMR effect in AFM materials, which largely facilitate the AFM memory device applications. These studies have excited a surge of attention on the AFM spintronics.[96,98–105]

At the same time, another major breakthrough in AFM spintronics came from the discovery of the Berry-curvature-induced anomalous Hall effect (AHE). In 2014, Chen *et al*.[106] first predicted the existence of the large AHE in cubic noncollinear antiferromagnets $Mn_3Ir$ due to the interband mixing of Bloch electrons induced by electric fields (**Figure 8**a & b), which is directly related to the topological features of Bloch bands and can be expressed as the integration over the Fermi sea of the Berry curvature of each occupied band.[107–110] Later in the same year, Kübler and Felser theoretically predicted the AHE in hexagonal $Mn_3Sn$ and $Mn_3Ge$.[111] Shortly thereafter, in 2015, Nakatsuji *et al*. experimentally observed the large AHE in bulk $Mn_3Sn$ single crystals (**Figure 8**c & d).[112] These results are rather exciting because they counter the long-standing wisdom that the AHE is an intrinsic feature of FM materials. The AHE in AFM materials has also been revealed in other systems.[113–115]

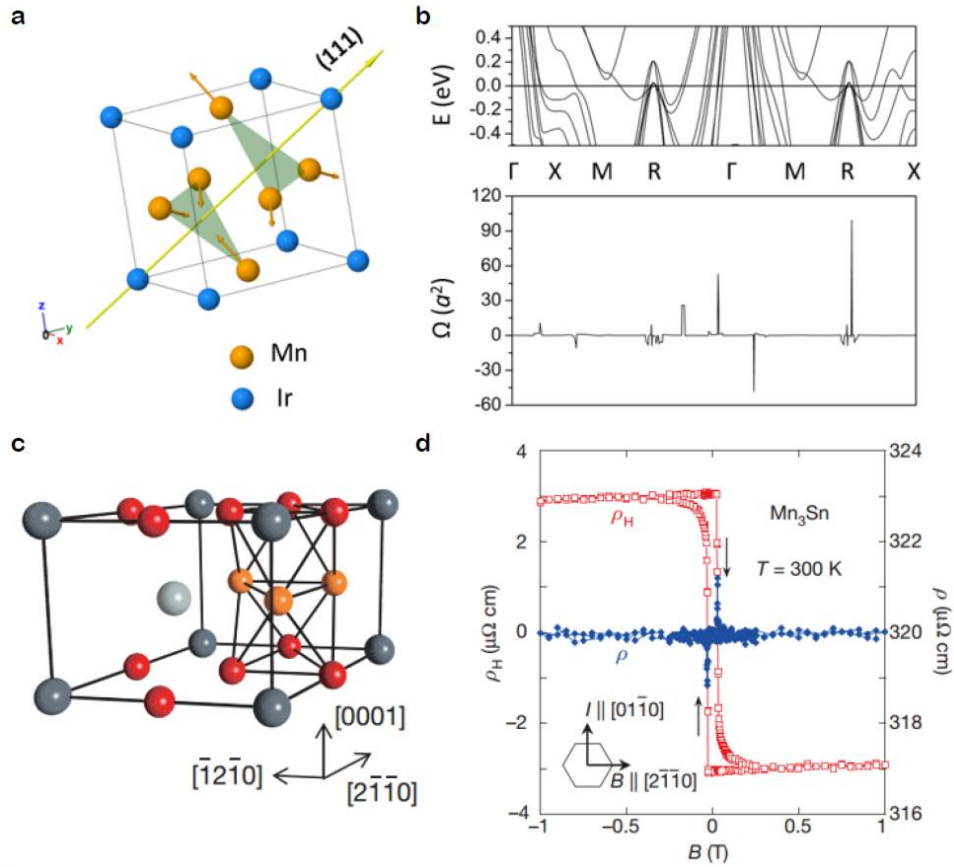

**Figure 8.** Berry curvature induced AHE in noncollinear antiferromagnets. a) Schematic of the crystal and magnetic structures of $Mn_3Ir$. b) Band structure (upper panel) near the Fermi level and Berry curvature (lower panel) along the same *k*-point path as the upper panel of the $Mn_3Ir$. Reproduced with permission.[106] Copyright 2014, American Physical Society. c) Schematic of the $Mn_3Sn$ crystal structure. d) Transverse and longitudinal resistivity as a function of the magnetic field of $Mn_3Sn$ at room temperature. Reproduced with permission.[112] Copyright 2015, Springer Nature.

Recently, the concept of topological AFM spintronics has been emphasized.[104,116,117] Basically, it is an emerging research field that especially focuses on the links between AFM spintronics and topological structures in real and momentum space, such as Majorana fermions in AFM topological superconductors,[104] topologically protected AFM skyrmions,[118–124] exotic AHE[125–127] and magnetic Weyl fermions[128–131] in AFM systems. The intimate relations between AFM spintronics and the current interest in topology in this cutting-edge research area could be powerful in creating fascinating and unprecedented opportunities for next generation information

## 6. Electric-field control of the AFM order parameter

Mn$_3$Pt, a cubic AFM intermetallic alloy, has a similar crystal structure as noncollinear AFM Mn$_3$Ir, which has been predicted to exhibit a large AHE.[106] More interestingly, it is a noncollinear antiferromagnet below ~365 K (**Figure 9**a),[132–134] while it becomes a collinear antiferromagnet (**Figure 9**b) through a first-order AFM-AFM transformation while heating up. At the same time, the lattice constant $a$ has an abrupt increase of ~0.8%.[133] Thus, the AFM spin structure of Mn$_3$Pt is intimately related to the lattice degree of freedom, which highly resembles FeRh, as stated before.

Similar to the work that reports the electric-field control of the magnetic order in FeRh,[69] high-quality Mn$_3$Pt films were epitaxially grown on FE BTO substrates, and the AHE was examined.[135] It was found that a large AHE exists in the low-temperature noncollinear AFM phase, while it is absent in the high-temperature collinear AFM phase (**Figure 9**d). More importantly, an in-plane compressive piezoelectric strain of ~0.3% enhances the AFM-AFM transformation temperature by ~25 K. Around the first-order MPT temperature, the electric-field-generated piezoelectric strain (**Figure 9**c) from the BTO crystals reversibly converts the spin structure between the noncollinear and collinear phases, consequently resulting in electrical switching of the AHE in Mn$_3$Pt (**Figure 9**e).

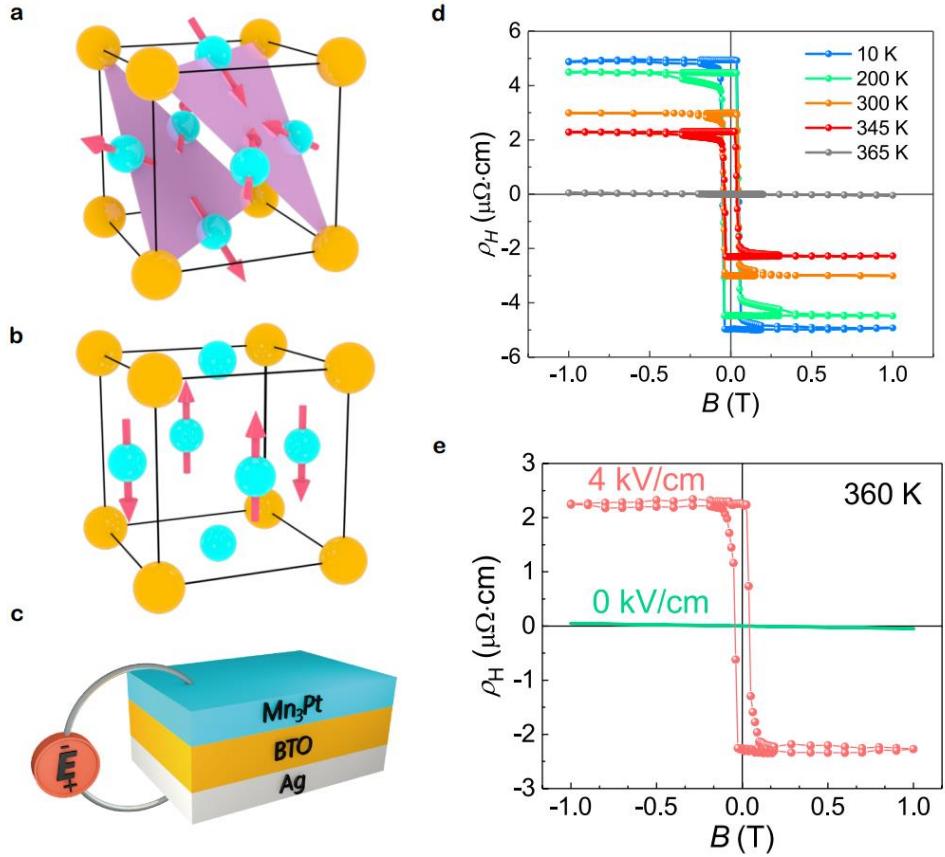

**Figure 9.** Electrical switching of the AHE in $Mn_3Pt$. a,b) Schematics of the crystal and magnetic structures of the low-temperature noncollinear and the high-temperature collinear phases of $Mn_3Pt$, respectively. c) Schematic of a 20 nm $Mn_3Pt$/BTO heterostructure with an electric field $E$ applied perpendicular to the BTO substrate. d) Hall resistivity versus magnetic field at different temperatures. e) Hall effect of the $Mn_3Pt$ film under zero electric field and $E$ = 4 kV cm$^{-1}$ at 360 K.[136]

This work could be important for the development of topological AFM spintronics because the electric-field-controlled devices are much less power consuming compared with other AFM spintronic devices operated by magnetic fields[66] or electrical currents.[97] In addition, it opens a new avenue to manipulating the AFM order parameter through electric fields.[136–138]

## 7. What is beyond?

Although giant magnetization and resistivity modulation have been achieved by small electric fields, such modulation could work most effectively around the phase transition temperature of

FeRh due to the phase instability around the phase transition.[69] However, the phase transition temperature of equiatomic FeRh is relatively high, ~360 K. First, it is necessary to dope FeRh with Pd, Ga or Ni to shift the transition temperature to be close to ~300 K. Second, it would be rather attractive to further amplify the resistance change of FeRh via the electric-field-controlled spin valve/magnetic tunnel junction structure, as schematized in **Figure 10**a & b. The reason is that the perpendicular junction geometry is capable of both larger resistance change ratios and higher-density memory device integration. Third, the recent study on the high-quality growth of FeRh films on MgO substrates by molecular beam epitaxy has achieved a large resistivity difference of ~80% between the AFM and FM phases of FeRh.[139] It is therefore possible to utilize MgO as a buffer layer material and the molecular beam epitaxy technique for further optimizing the crystalline quality of FeRh films while they are grown on FE substrates, which would yield a larger electroresistance modulation to facilitate device applications.

On the other hand, when the AFM spintronics marries the topology that is of intense interest in the current condensed matter physics, topological AFM spintronics is becoming a center of focus in the field of magnetism and magnetic materials. For example, experimental evidence for time-reversal-symmetry-breaking Weyl fermions has been reported by Kuroda *et al*. in AFM $Mn_3Sn$,[128] where both Weyl points near the Fermi level (**Figure 10**c & d) and the chiral anomaly have been demonstrated via the angle-resolved photoemission spectroscopy and magnetotransport measurements, respectively. Basically, Weyl points are topological objects that appear in pairs with opposite chiralities and act as magnetic monopoles (or sources of the Berry flux) in momentum space, which are usually the crossing points when two nondegenerate bands cross linearly.[140] A material system with Weyl points must have broken time reversal symmetry or inversion symmetry. Earlier experimental realization of Weyl fermions[141,142] in nonmagnetic material systems were all based on inversion symmetry breaking, and thus, Kuroda et al.'s work[128] is the first experimental report toward time-reversal-symmetry-breaking Weyl fermions in spite of the many theoretical predictions.[140,143–148]

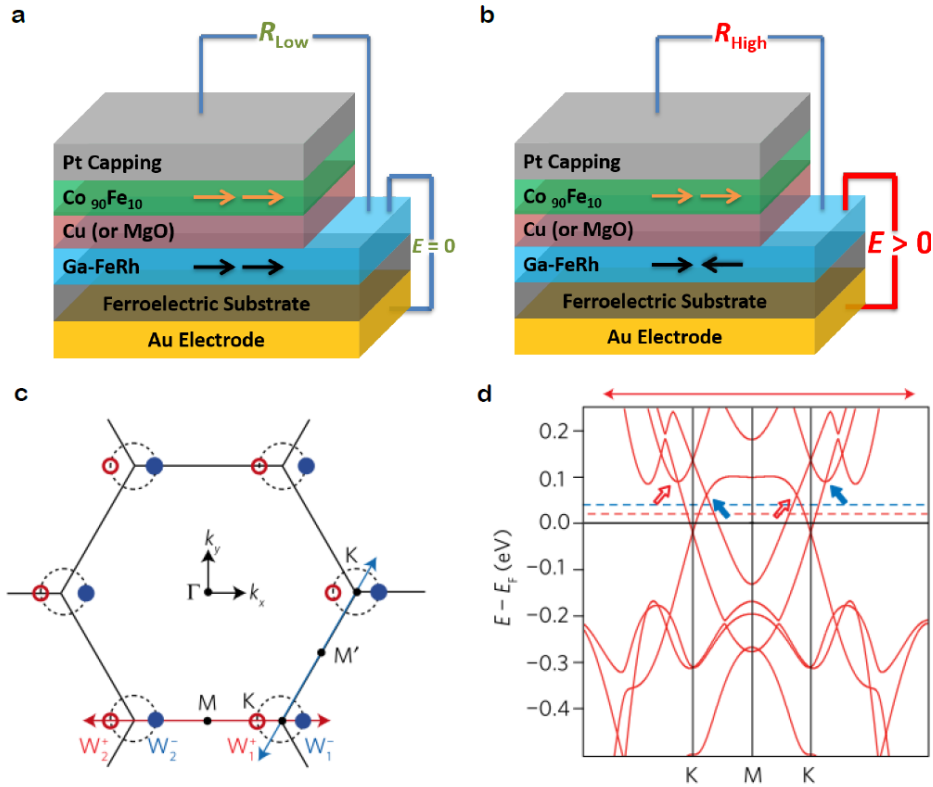

**Figure 10.** Future directions. a,b) Schematics of low and high resistance states of electric-field-controlled FeRh-based spin valve/magnetic tunnel junction structure operated at room temperature. Theoretically calculated band structure of $Mn_3Sn$. c) Distribution of the Weyl points in the bands on the $k_x$–$k_y$ plane at $k_z = 0$ near $E_F$ for the magnetic texture of noncollinear $Mn_3Sn$. Two pairs of Weyl nodes with different chirality ($W^+$, $W^-$) are shown by open and solid circles. The dotted circles schematically show the hypothetical nodal rings that appear when SOC is turned off. d) Enlarged DFT band structure around the M point cut along distinct high-symmetry lines, which correspond to the red arrows in (c). Weyl (band crossing) points with opposite chirality are denoted by open and solid arrows. Reproduced with permission.[128] Copyright 2017, Springer Nature.

Since the wavelengths of the Bloch waves of electrons in solid-state materials are comparable with the lattice constants of the crystals, the electronic properties of the solid-state materials based on the band structures of the electrons are rather sensitive to the lattice modulation, which in turn can be conveniently achieved by electric fields via functional FE materials. Therefore, it is possible to electrically control the magnetic Weyl points in $Mn_3Sn$ via this approach once high-quality thin films can be fabricated onto FE substrates. In addition, this approach could also be useful to tune the topological protection and Dirac points by manipulating the AFM order parameters.[104] Overall, using electric fields rather than magnetic fields or electrical currents to control the electronic properties of magnetic materials is of great potential for low energy-consuming information device applications, which is even powerful for tailoring exotic physical phenomena in the emerging topological AFM spintronics and thus paves the way for realizing novel low-power electronic device-based topological effects.

**Acknowledgements**
Z.Q.L. acknowledges financial support from the National Natural Science Foundation of China (NSFC Grant No. 51822101, 51861135104, 51771009)

Received: ((will be filled in by the editorial staff))
Revised: ((will be filled in by the editorial staff))
Published online: ((will be filled in by the editorial staff))

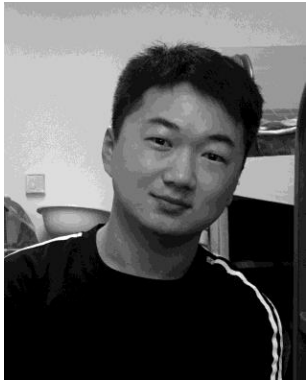

Mr. Zexin Feng obtained his BoS degree from Beihang University and is now a graduate student in the Functional Thin Film Lab of Beihang University led by Prof. Zhiqi Liu.

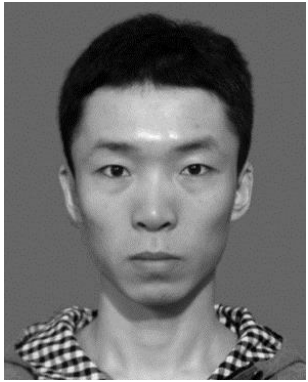

Mr. Han Yan obtained his BoS degree from Wuhan University of Technology and is now a graduate student in the Functional Thin Film Lab of Beihang University led by Prof. Zhiqi Liu.

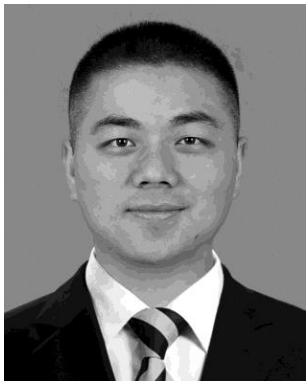

Prof. Zhiqi Liu obtained his BoS degree from Lanzhou University and PhD degree from National University of Singapore. Afterward, he performed postdoc research at Oak Ridge National Laboratory, University of California, Berkeley and Los Alamos National Laboratory. He is now a faculty professor at School of Materials Science and Engineering of Beihang University and the director of the Functional Thin film Lab there.